\documentclass[aps,prl,reprint,superscriptaddress]{revtex4-1}

\usepackage{graphicx,topcapt,booktabs,SIunits,commath,units,mhchem,amsmath}

\begin{document}

\title{Key $^{19}$Ne states identified affecting $\gamma$-ray emission from $^{18}$F in novae}

\author{M.R.~Hall}
\email{mhall12@alumni.nd.edu}
\author{D.W.~Bardayan}
\affiliation{Department of Physics, University of Notre Dame, Notre Dame, Indiana 46556, USA}
\author{T.~Baugher}
\affiliation{Department of Physics and Astronomy, Rutgers University, New Brunswick, New Jersey 08903, USA}
\author{A.~Lepailleur}
\affiliation{Department of Physics and Astronomy, Rutgers University, New Brunswick, New Jersey 08903, USA}
\author{S.D.~Pain}
\affiliation{Physics Division, Oak Ridge National Laboratory, Oak Ridge, Tennessee 37831, USA}
\author{A.~Ratkiewicz}
\affiliation{Department of Physics and Astronomy, Rutgers University, New Brunswick, New Jersey 08903, USA}

\author{S.~Ahn}
\affiliation{National Superconducting Cyclotron Laboratory, Michigan State University, East Lansing, Michigan 48824, USA}

\author{J.M.~Allen}
\affiliation{Department of Physics, University of Notre Dame, Notre Dame, Indiana 46556, USA}

\author{J.T.~Anderson}
\affiliation{Physics Division, Argonne National Laboratory, Argonne, Illinois 60439, USA}

\author{A.D.~Ayangeakaa}
\affiliation{Physics Division, Argonne National Laboratory, Argonne, Illinois 60439, USA}

\author{J.C.~Blackmon}
\affiliation{Department of Physics and Astronomy, Louisiana State University, Baton Rouge, Louisiana 70803, USA}

\author{S.~Burcher}
\affiliation{Department of Physics and Astronomy, University of Tennessee, Knoxville, Tennessee 37996, USA}

\author{M.P.~Carpenter}
\affiliation{Physics Division, Argonne National Laboratory, Argonne, Illinois 60439, USA}

\author{S.M.~Cha}
\affiliation{Department of Physics, Sungkyunkwan University, Suwon 16419, South Korea}

\author{K.Y.~Chae}
\affiliation{Department of Physics, Sungkyunkwan University, Suwon 16419, South Korea}

\author{K.A.~Chipps}
\affiliation{Physics Division, Oak Ridge National Laboratory, Oak Ridge, Tennessee 37831, USA}

\author{J.A.~Cizewski}
\affiliation{Department of Physics and Astronomy, Rutgers University, New Brunswick, New Jersey 08903, USA}

\author{M.~Febbraro}
\affiliation{Physics Division, Oak Ridge National Laboratory, Oak Ridge, Tennessee 37831, USA}

\author{O.~Hall}
\affiliation{Department of Physics, University of Notre Dame, Notre Dame, Indiana 46556, USA}
\affiliation{Department of Physics, University of Surrey, Guildford, Surrey GU2 7XH, United Kingdom}

\author{J.~Hu}
\affiliation{Department of Physics, University of Notre Dame, Notre Dame, Indiana 46556, USA}

\author{C.L.~Jiang}
\affiliation{Physics Division, Argonne National Laboratory, Argonne, Illinois 60439, USA}

\author{K.L.~Jones}
\affiliation{Department of Physics and Astronomy, University of Tennessee, Knoxville, Tennessee 37996, USA}

\author{E.J.~Lee}
\affiliation{Department of Physics, Sungkyunkwan University, Suwon 16419, South Korea}

\author{P.D.~O'Malley}
\affiliation{Department of Physics, University of Notre Dame, Notre Dame, Indiana 46556, USA}

\author{S.~Ota}
\affiliation{Physics Division, Lawrence Livermore National Laboratory, Livermore, California 94551, USA}

\author{B.C.~Rasco}
\affiliation{Department of Physics and Astronomy, Louisiana State University, Baton Rouge, Louisiana 70803, USA}

\author{D.~Santiago-Gonzalez}
\affiliation{Department of Physics and Astronomy, Louisiana State University, Baton Rouge, Louisiana 70803, USA}

\author{D.~Seweryniak}
\affiliation{Physics Division, Argonne National Laboratory, Argonne, Illinois 60439, USA}

\author{H.~Sims}
\affiliation{Department of Physics and Astronomy, Rutgers University, New Brunswick, New Jersey 08903, USA}
\affiliation{Department of Physics, University of Surrey, Guildford, Surrey GU2 7XH, United Kingdom}

\author{K.~Smith}
\affiliation{Department of Physics and Astronomy, University of Tennessee, Knoxville, Tennessee 37996, USA}

\author{W.P.~Tan}
\affiliation{Department of Physics, University of Notre Dame, Notre Dame, Indiana 46556, USA}

\author{P.~Thompson}
\affiliation{Physics Division, Oak Ridge National Laboratory, Oak Ridge, Tennessee 37831, USA}
\affiliation{Department of Physics and Astronomy, University of Tennessee, Knoxville, Tennessee 37996, USA}

\author{C.~Thornsberry}
\affiliation{Department of Physics and Astronomy, University of Tennessee, Knoxville, Tennessee 37996, USA}

\author{R.L.~Varner}
\affiliation{Physics Division, Oak Ridge National Laboratory, Oak Ridge, Tennessee 37831, USA}

\author{D.~Walter}
\affiliation{Department of Physics and Astronomy, Rutgers University, New Brunswick, New Jersey 08903, USA}

\author{G.L.~Wilson}
\affiliation{Department of Physics and Astronomy, Louisiana State University, Baton Rouge, Louisiana 70803, USA}
\affiliation{Department of Physics and Applied Physics, University of Massachusetts Lowell, Lowell, Massachusetts 01854, USA}

\author{S.~Zhu}
\affiliation{Physics Division, Argonne National Laboratory, Argonne, Illinois 60439, USA}

\date{\today}

\begin{abstract}

Detection of nuclear-decay $\gamma$ rays provides a sensitive thermometer of nova nucleosynthesis. The most intense $\gamma$-ray flux is thought to be annihilation radiation from the $\beta^+$ decay of $^{18}$F, which is destroyed prior to decay by the $^{18}$F($p$,$\alpha$)$^{15}$O reaction. Estimates of $^{18}$F production had been uncertain, however, because key near-threshold levels in the compound nucleus, $^{19}$Ne, had yet to be identified. This Letter reports the first measurement of the $^{19}$F($^{3}$He,$t\gamma$)$^{19}$Ne reaction, in which the placement of two long-sought 3/2$^+$ levels is suggested via triton-$\gamma$-$\gamma$ coincidences. The precise determination of their resonance energies reduces the upper limit of the rate by a factor of $1.5-17$ at nova temperatures and reduces the average uncertainty on the nova detection probability by a factor of 2.1.

\end{abstract}

\maketitle{}

The outburst of energy that occurs once a white dwarf accretes a sufficient amount of material from a less-evolved companion star is called a classical nova. Novae are fairly common events in the Milky Way, with $\sim$50 estimated to occur per year \cite{Shafter2017}. While many unstable isotopes are created by novae during the hot carbon-nitrogen-oxygen cycles, few have been postulated to produce detectable $\gamma$ rays in the~keV-MeV energy range \cite{Jose2007}. The largest $\gamma$-ray flux from novae is predicted to be from energies $\le511$~keV, due to the annihilation of positrons created from the $\beta^+$ decays of $^{13}$N and $^{18}$F \cite{Hernanz2006}. The main contributor to the flux of the annihilation $\gamma$ rays has been identified as $^{18}$F because its half-life ($t_{1/2} = 109.77(5)$ min \cite{Tilley1995}) allows it to survive until the envelope of the explosion becomes transparent to $\gamma$ rays. Detection of this radiation would provide a test of nova models, which currently fail to reproduce observed properties such as the total ejected mass  \cite{Jose2008}.

Reliable estimates of the sensitivity required for detection have been impossible to determine. This is because the destruction of $^{18}$F prior to its $\beta^+$ decay, which occurs primarily via the $^{18}$F($p$,$\alpha$)$^{15}$O reaction, was not sufficiently known. The $^{18}$F($p$,$\alpha$)$^{15}$O reaction-rate uncertainty at nova temperatures ($T=0.1-0.4$ GK) is attributed to the unknown energies of, and interference between $s$-wave ($J^\pi = 1/2^+, 3/2^+$) resonances corresponding to states of the same spin in the compound nucleus, $^{19}$Ne \cite{Dufour2007, Beer2011}. Precise determination of these resonances would greatly reduce the reaction-rate uncertainty.

\begin{figure}[ht!]
\begin{center}
\includegraphics[width=\linewidth]{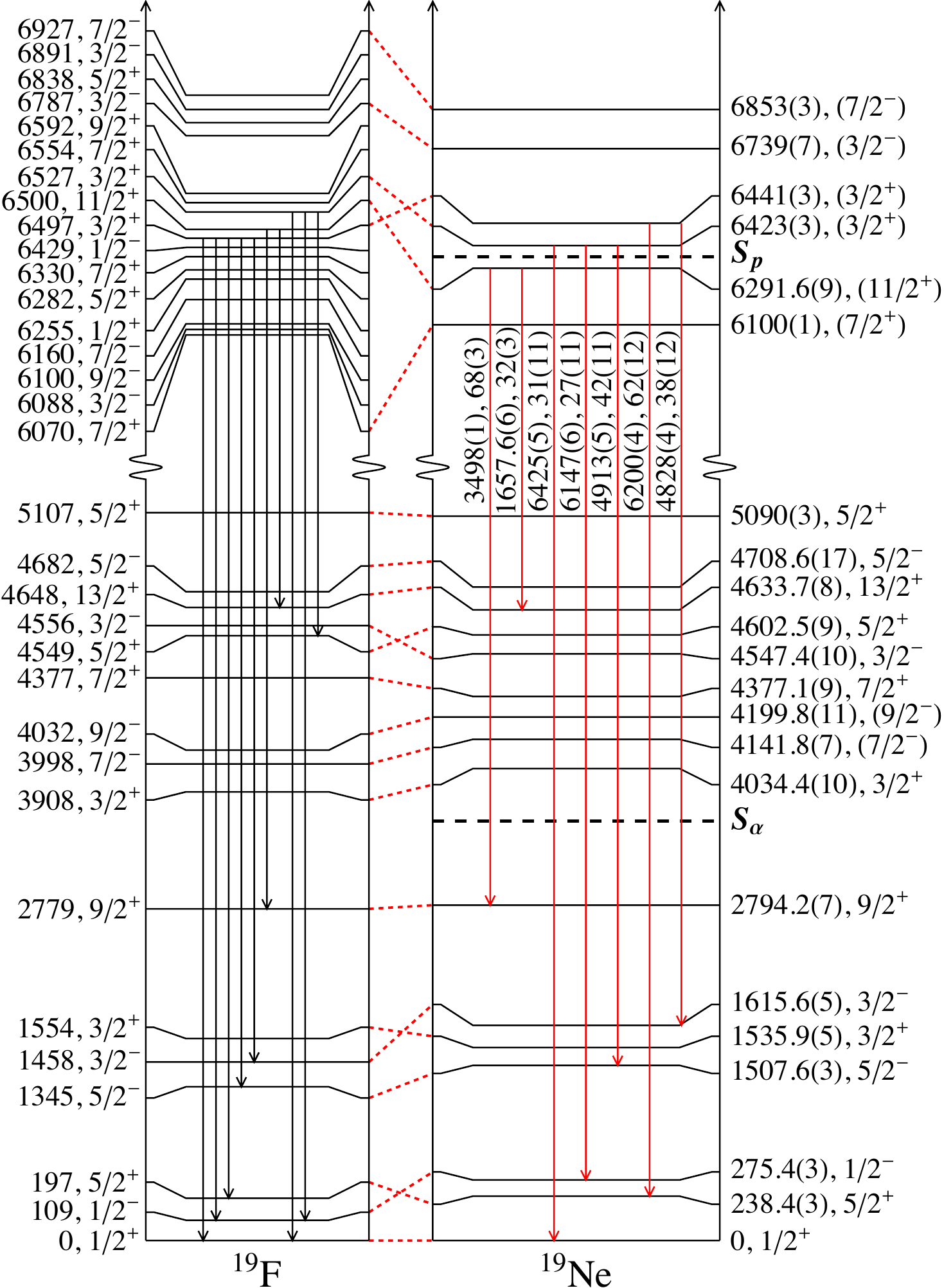}

\caption{(Color online) Level structures of $^{19}$F and $^{19}$Ne. The $^{19}$Ne $\gamma$-ray energies (keV), level energies (keV), and branching ratios (\%) were determined from this experiment. Mirror connections between $^{19}$Ne and $^{19}$F states are shown with dashed lines. Plotted $^{19}$Ne transitions were first observed in this work, and $^{19}$F transitions \cite{Tilley1995} referenced in the text are shown. $^{19}$F and $^{19}$Ne levels between 5100 and 6000~keV are omitted for clarity. }

\label{Lev}
\end{center}
\end{figure}

Based on mirror symmetry, there should be two near-threshold 3/2$^+$ states in $^{19}$Ne, corresponding to the 6497- and 6527-keV states in $^{19}$F \cite{Tilley1995}. The cross section exhibits interference between these states and a broad 3/2$^+$ resonance at $E_{cm}=665$~keV \cite{Bardayan2001}. This interference is a dominant source of uncertainty in the reaction rate \cite{Beer2011, Laird, Bardayan2015} and strongly depends on the energies and proton widths of the ``missing" 3/2$^+$ $^{19}$Ne levels.  A compilation by Nesaraja \textit{et al.} \cite{Nesaraja2007} estimated the average energy shift from states in the mirror nucleus above 6400~keV to be $50\pm30$~keV, though the actual shift for individual levels could be larger. This means the two 3/2$^+$ states should have energies of $6447\pm30$~keV and $6477\pm30$~keV in $^{19}$Ne. Estimates of the reaction rate, including the uncertainties for these energy levels and their interference with the 665-keV state, results in the $^{18}$F($p$,$\alpha$)$^{15}$O rate being uncertain by factors between 3 and 33 at temperatures of $0.1-0.25$~GK, the most important range for nova nucleosynthesis. 

Because of their importance, a number of experiments have searched for the $J^\pi=3/2^+$ levels above the proton  ($^{18}$F+$p$) threshold ($S_p = 6410$~keV) \cite{Utku1998, Adekola2011, Laird, Parikh2015, Bardayan2015}. Utku \textit{et al.} \cite{Utku1998} showed the presence of potential resonances at 8 and 38~keV using the $^{19}$F($^3$He,$t$)$^{19}$Ne reaction and explored a $3/2^+$ spin-parity assignment to both. Adekola \textit{et al.} \cite{Adekola2011} reconstructed neutron angular distributions from the $^{18}$F($d$,$n$)$^{19}$Ne reaction but were only able to set upper limits on the strength of the 3/2$^+$ levels. Laird {\it et al.} \cite{Laird} deconvolved the counts in a near-threshold triplet to conclude none seemed consistent with a $3/2^+$ assignment based on measured angular distributions. Kozub \textit{et al.} \cite{Kozub2005} and de~S\'er\'eville \textit{et al.} \cite{Sereville2005, Sereville2007} observed a strong 3/2$^+$ level population in $^{18}$F($d$,$p$)$^{19}$F measurements, but could not accurately estimate its placement in the mirror, $^{19}$Ne.

All previous measurements relied upon charged-particle spectroscopy and thus were limited by the energy resolution at which these particles could be detected. This Letter details the first detection of $\gamma$ rays from the de-excitation of these closely-spaced, near-threshold levels in $^{19}$Ne, instead of solely relying on charged-particle detection. This was not previously attempted due to the small $\gamma$-ray branching ratios ($10^{-3}$ \cite{Nesaraja2007}) expected for energy levels above the proton threshold.  

To search for the 3/2$^+$ states of interest, the $^{19}$F($^{3}$He,$t\gamma$)$^{19}$Ne reaction was studied using GODDESS (Gammasphere ORRUBA Dual Detectors for Experimental Structure Studies) \cite{Rat2013, Pain2014, Pain2017} at Argonne National Laboratory. A beam of 30-MeV $^{3}$He from the ATLAS accelerator impinged on a 938-$\mu$g/cm$^2$ CaF$_2$ target with an average beam intensity of 2.5~pnA. The reaction tritons were measured over laboratory angles of $18^\circ-90^\circ$ in the silicon detector array ORRUBA (Oak Ridge Rutgers University Barrel Array) \cite{Pain2007}, which was augmented by custom endcap silicon detector $\Delta E$-$E$ telescopes  \cite{Pain2017}. Coincident $\gamma$ rays were detected using the Compton-suppressed high-purity germanium detector array Gammasphere \cite{Lee1992}. A 0.5-mm thick aluminum plate, which was thin enough to transmit the tritons of interest, was mounted in front of the endcap telescopes to suppress the high rate of elastically-scattered $^3$He.

Calibrations of the Gammasphere detectors were performed with sources of $^{152}$Eu, $^{56}$Co, and $^{238}$Pu+$^{13}$C, covering an energy range of 122 to 6128~keV. Systematic uncertainties in the $\gamma$-ray energy calibration ($\sim 0.3-2.0$~keV) were combined in quadrature with the statistical errors in the peak centroids to calculate the reported energy uncertainties. 

The $^{19}$Ne energy levels reconstructed from the detected $\gamma$ rays were highly constrained by the gates placed on the data. With the exception of transitions directly to the ground state, triton-$\gamma$-$\gamma$ coincidences were used to identify the transitions. For levels that decayed through multiple $\gamma$-ray cascades, the excitation energies were determined by averaging the summed level energies for each individual $\gamma$-ray cascade, weighted by their uncertainty. In addition, much of the $\gamma$-ray background was removed by requiring a tight time coincidence between Gammasphere and ORRUBA, and the remaining random background was characterized by gating adjacent to the timing peak.

Above 1000 keV, only the 1508- and 4634-keV states have lifetimes long enough ($\tau \approx 1.7$ ps and $\tau>1.0$ ps, respectively  \cite{Tan2005}) to allow the $^{19}$Ne to stop in the target before de-excitation. In all other cases, the $^{19}$Ne will still be travelling when $\gamma$ decay occurs, and therefore, the $\gamma$-ray spectra needed to be Doppler corrected. While small, values of $\beta$ ranged between 0.005 and 0.025 and were calculated on an event-by-event basis from the detected triton energy and angle. The sharpest $\gamma$-ray peaks were obtained assuming the recoil $^{19}$Ne ions lost no energy before decaying.

In total, 41 decays from 21 energy levels were identified \cite{Hall2019}, including seven decays from three near-threshold levels of astrophysical interest. Figure \ref{Lev} displays the $^{19}$Ne level scheme, reconstructed from decays observed in the data, next to the $^{19}$F level scheme. Newly observed transitions from $^{19}$Ne states around 6400~keV and previously observed $^{19}$F transitions for their proposed mirrors are shown, with $\gamma$-ray energies and branching ratios appearing next to each arrow. Results from the near-threshold levels are highlighted in the following discussions. 

A strong subthreshold state was found at 6289~keV by Adekola \textit{et al.} \cite{Adekola2011}, which was later shown to be a doublet by Parikh \textit{et al.} \cite{Parikh2015}. Bardayan \textit{et al.} \cite{Bardayan2015} determined the low-spin member to be 1/2$^+$, while Laird \textit{et al.} \cite{Laird} showed the other member has high spin. Two strong transitions were observed to the 4634- (13/2$^+$) and 2794-keV (9/2$^+$) levels, and averaging the summed cascade energies results in a best value of $E_x=6291.7\pm0.9$~keV for the level energy. A comparison with the known $^{19}$F energy levels and associated $\gamma$ decays between 6000 and 7000~keV suggests the spin-parity of this state is 11/2$^+$, making it the mirror of the $E_x(^{19}\mathrm{F}) = 6500$~keV level. No $\gamma$ decays from the 1/2$^+$ state were observed in this work, which is not surprising, since none have been observed for the mirror state in $^{19}$F at $E_x = 6255$~keV \cite{Tilley1995}.

\begin{figure}[h!]
\begin{center}
\includegraphics[width=\linewidth]{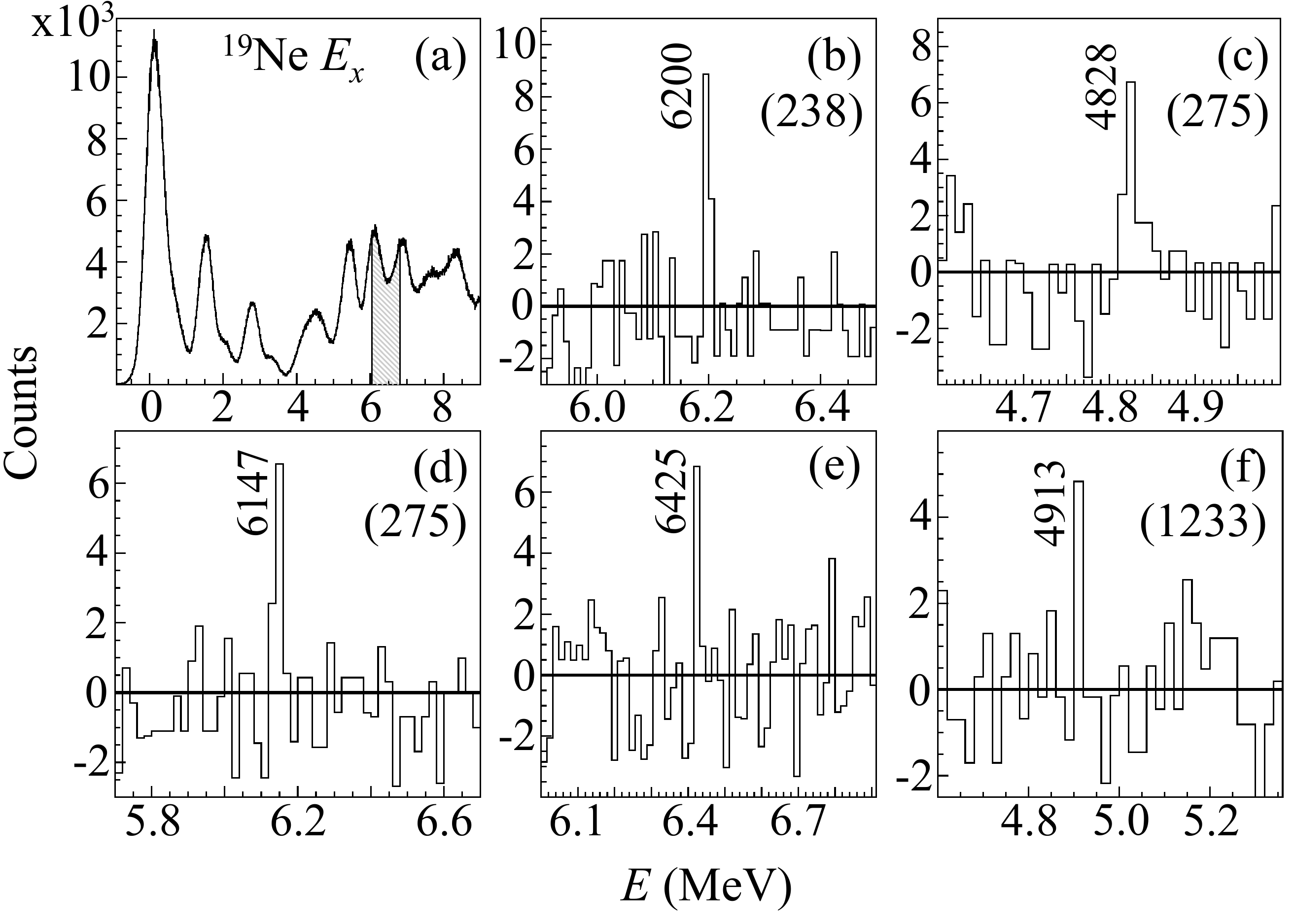}

\caption{(a) $^{19}$Ne excitation-energy spectrum reconstructed from the detected tritons at $\theta_{lab} = 20^\circ$. (b-f) Random-subtracted $\gamma$-ray spectra (10, 10, 20, 16, and 20 keV/bin, respectively) from the two lowest above-threshold states. The spectra were gated on the shaded excitation-energy region in (a) (6.0-6.6 MeV) and the $\gamma$ ray shown in parentheses, with the exception of (e), which is a ground state transition. }
\label{spectra}
\end{center}
\end{figure}

Above the proton threshold, the next grouping of levels shown to be populated by the $^{19}$F($^{3}$He,$t\gamma$)$^{19}$Ne reaction \cite{Utku1998, Laird, Parikh2015} has been a source of debate due to their potentially important contributions to the reaction rate. The present data show transitions to the ground state (1/2$^+$), 275-keV (1/2$^-$) and 1508-keV (5/2$^-$) states from an energy level at $6423\pm3$~keV ($E_{cm}=13$ keV). Additionally, decays to the 238-keV (5/2$^+$) and 1616-keV (3/2$^-$) states were observed from a level with excitation energy $6441\pm3$~keV ($E_{cm}=31$ keV). The average uncertainties on the $\gamma$-ray peak centroids and $\gamma$-ray energy calibration were $\approx$3 keV and $\approx$1 keV, respectively. Figure \ref{spectra} shows the $^{19}$Ne excitation energy ($E_x$) spectrum generated from the detected reaction tritons and all five of the $\gamma$-ray peaks mentioned above. The low spins of the levels decayed to and previous discussion of the 6292-keV state contradict the assertion by Laird \textit{et al.} \cite{Laird} that the 6440-keV state is the mirror of the $E_x=6500$-keV $^{19}$F state~and~has~an~11/2$^+$~spin-parity.

The most likely spin assignment for both the 6423- and 6441-keV states is 3/2$^+$ based on the known  levels and $\gamma$-ray transitions from the $^{19}$F mirror states. The 6497-keV state in $^{19}$F decays to the mirrors of the ground state, 238-, 275-, 1508-, and 1616-keV $^{19}$Ne states, whereas the 6527-keV state in $^{19}$F decays to the mirrors of the ground state, 275-, and 4603-keV $^{19}$Ne states \cite{Tilley1995}. Two of three decays from the 6423-keV state have been previously observed from the $E_x(^{19}\mathrm{F})=6527$-keV level, whereas both decays from the 6441-keV state have been observed from the $E_x(^{19}\mathrm{F})=6497$-keV state. Therefore, mirror connections between the 6497-keV $^{19}$F and 6441-keV $^{19}$Ne states and between the 6527-keV $^{19}$F and 6423-keV $^{19}$Ne states are suggested. The only other possible spin-parity for these states consistent with the energy levels of the mirror and the multipolarity of the transitions is 7/2$^+$. However, the decay scheme for the 7/2$^+$ mirror is quite different than what was observed, and thus this seems unlikely. In any case, such a level would have limited importance to the $^{18}$F($p$,$\alpha$)$^{15}$O rate because of the lack of interference with any broad resonance.

These two 3/2$^+$ states near $E_x=6400$~keV would have been observable by Adekola \textit{et al.} \cite{Adekola2011} using the $^{18}$F($d$,$n$)$^{19}$Ne reaction if the states were of sufficient strength and could be resolved from other states. Upper limits for the spectroscopic factor  ($S_p\le0.028$) and proton width ($\Gamma_p\le2.35\times10^{-15}$~keV) were set for a $3/2^+$ state in this excitation-energy region \cite{Adekola2011,Adekola2011_2}. To be consistent, the following calculations assume most of the spectroscopic strength to be in one of the two 3/2$^+$ states (which was observed in $^{18}$F($d$,$p$)$^{19}$F measurements \cite{Kozub2005, Sereville2007}) and scale the widths with energy accordingly. The mirror assignments for the candidate $3/2^+$ states could be reversed, but this would not affect the results since the widths were determined in previous experiments and not derived from those states in $^{19}$F. 

\begin{figure}[h!]
\begin{center}
\includegraphics[width=\linewidth]{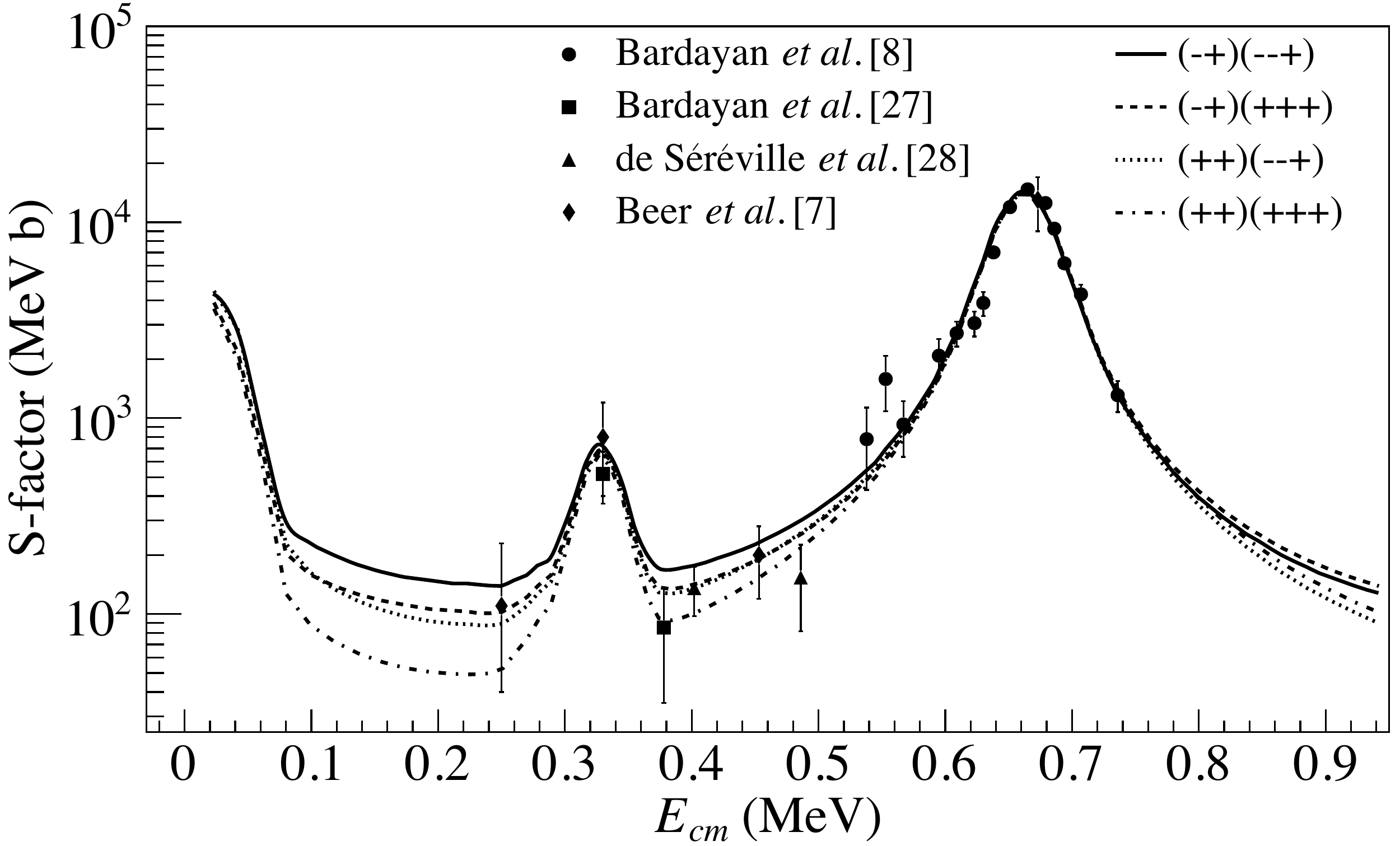}

\caption{$^{18}$F($p$,$\alpha$)$^{15}$O S-factors calculated with \textsc{Azure2}. The first and second sets of parentheses show the interference sign between the two 1/2$^+$ and three 3/2$^+$ states, respectively, in order of increasing energy. Intermediate S-factors from other interference patterns for the 3/2$^+$ states were omitted for clarity. Experimental data from Refs. \cite{Beer2011,Bardayan2001,Bardayan2002,Sereville2009} are included.
}
\label{Sfactor}
\end{center}
\end{figure}

To assess the $^{18}$F($p$,$\alpha$)$^{15}$O rate uncertainties due to interference between $s$-wave resonances, the R-Matrix code \textsc{Azure2} \cite{Azuma2010} was used. The reaction rate is calculated from the astrophysical S-factor, which is the reaction cross section with the strong energy dependence due to the Coulomb barrier penetration removed. The S-factors calculated using \textsc{Azure2} for various interference combinations are shown in Fig. \ref{Sfactor}. The R-Matrix channel radius used was 5.2 fm, and a 15-keV energy resolution was included in the calculation to directly compare with the available experimental data \cite{Bardayan2002,Beer2011, Bardayan2001,Sereville2009}. Table \ref{SfactTable} shows the level energies and widths used in the calculation. Constructive and destructive interference between the known 1/2$^+$ states and candidate 3/2$^+$ states is denoted by the first and second set of parentheses, respectively. The majority of the S-factor uncertainty comes from the unknown interference sign of the 1/2$^+$ states. Nonetheless, interference between the broad 3/2$^+$ state at 665~keV and the two near-threshold 3/2$^+$ states exacerbates this uncertainty. 

\begin{table}[h!]
\caption{\label{SfactTable}Resonance parameters used in the S-factor calculation. The widths were scaled within the energy uncertainty range for the reaction-rate calculation.}
\begin{ruledtabular}
\begin{tabular}{ccccc}
$E_x$ (keV) & $E_r$ (keV) & $J^\pi$ & $\Gamma_p$ (keV) & $\Gamma_\alpha$ (keV) \\
\hline
6286(3)$^\mathrm{a}$ & -124 & 1/2$^+$ & 83.5$^{\mathrm{c}}$ & 11.6 \\

6416(4)$^\mathrm{b}$ & 6 & 3/2$^-$ & 4.7$\times 10^{-50}$ & 0.5 \\

6423(3) & 13 & 3/2$^+$ & $\leq$3.9$\times 10^{-29}$ & 1.2 \\

6439(3)$^\mathrm{a}$ & 29 & 1/2$^-$ & $\leq$3.8$\times 10^{-19}$ & 220 \\

6441(3) & 31 & 3/2$^+$ & $\leq$8.4$\times 10^{-18}$ & 1.3 \\

6459(5)$^\mathrm{b}$ & 49 & 5/2$^-$ & 8.4$\times 10^{-14}$ & 5.5 \\

6699(3)$^\mathrm{a}$ & 289 & 5/2$^+$ & 2.4$\times 10^{-5}$ & 1.2 \\

6742(2)$^\mathrm{a}$ & 332 & 3/2$^-$ & 2.22$\times 10^{-3}$ & 5.2 \\

7075(2)$^\mathrm{a}$ & 665 & 3/2$^+$ & 15.2 & 23.8 \\

7871(19)$^\mathrm{a}$ & 1461 & 1/2$^+$ & 55 & 347 \\

\end{tabular}
\end{ruledtabular}
\begin{flushleft}
$^\mathrm{a}$All level parameters taken from Bardayan \textit{et al.} \cite{Bardayan2015}.

$^\mathrm{b}$All level parameters taken from Laird \textit{et al.} \cite{Laird}.

$^\mathrm{c}$ANC (fm$^{1/2}$).
\end{flushleft}
\end{table}

\begin{figure}[h!]
\begin{center}
\includegraphics[width=\linewidth]{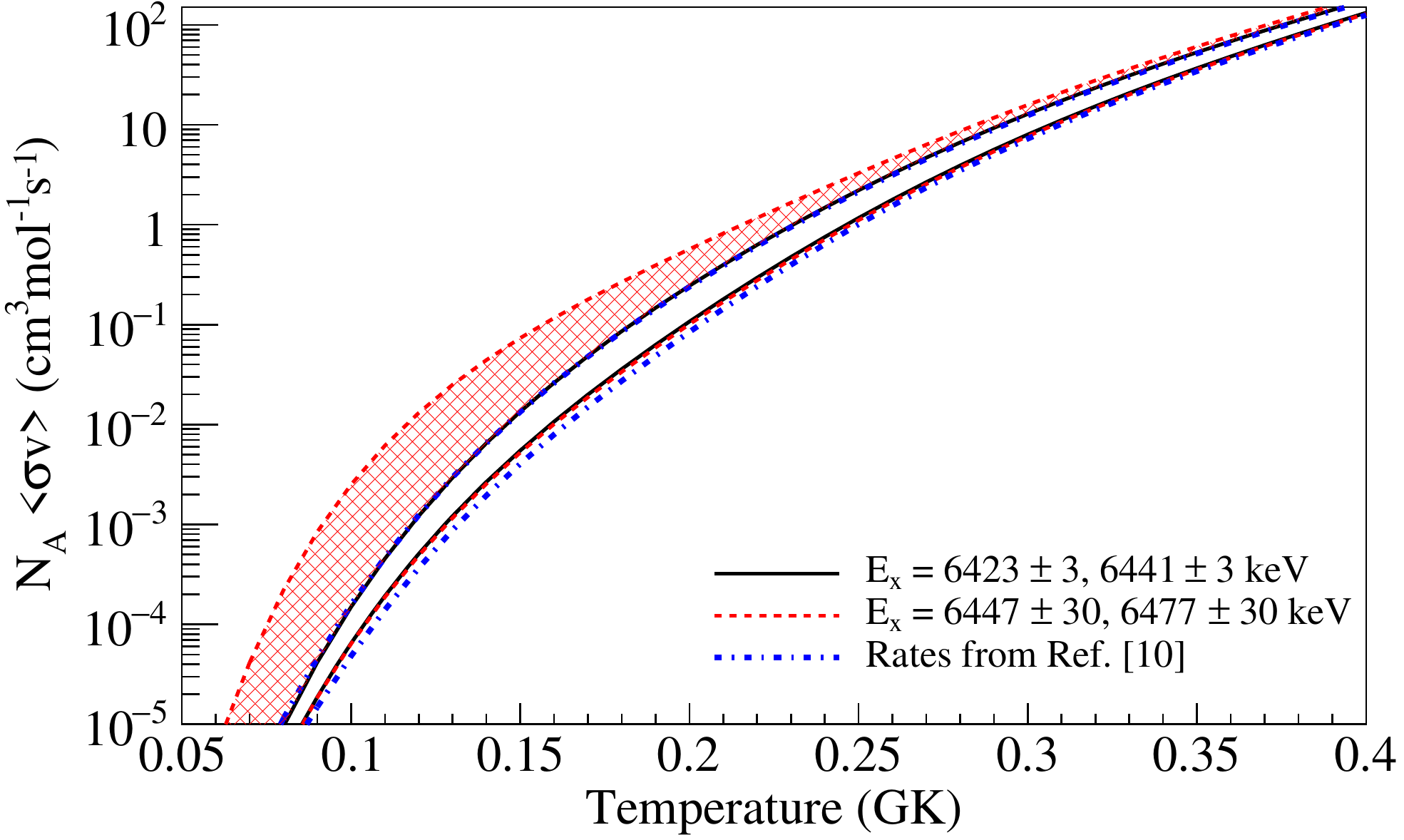}

\caption{(Color online) Reaction rates calculated using \textsc{Azure2}. The rate bands were produced by varying the level properties within uncertainty. The lower limits almost agree, but the upper limit is reduced by a factor of $1.5-17$ in the current work. The shaded region represents the values of the rate that are now excluded. The rates calculated by Ref. \cite{Bardayan2015} are shown for comparison (see text).}
\label{Rate}
\end{center}
\end{figure}

The effect of the interference between the three 3/2$^+$ states is better illustrated by calculating the rate using the range of excitation energies predicted for the 3/2$^+$ states. The energies and widths of the near-threshold levels were varied within uncertainty to calculate the upper and lower limits of the $^{18}$F($p$,$\alpha$)$^{15}$O reaction rate. This process was performed twice: first for the previous best estimates of $E_x=6447\pm30$~keV and $E_x=6477\pm30$~keV and then again using the newly-constrained values of $6423\pm3$~keV and $6441\pm3$~keV. Figure \ref{Rate} shows the $^{18}$F($p$,$\alpha$)$^{15}$O rate as a function of temperature, comparing the calculated upper and lower limits. Values for the rate calculated with proton widths less than the upper limit set by Adekola \textit{et al.} \cite{Adekola2011} fall within the rate bands.

Constraining the 3/2$^+$ states to $6423\pm3$~keV and $6441\pm3$~keV reduces the reaction-rate uncertainty at $T=0.25$~GK to 0.96~cm$^3$mol$^{-1}$s$^{-1}$, a reduction from the previous upper limit by a factor of 1.5, whereas at low temperatures ($T=0.1$~GK) the current uncertainty of $7.2\times10^{-5}$~cm$^3$mol$^{-1}$s$^{-1}$ represents a reduction from the previous upper limit by a factor of 17. For comparison, the previously accepted rate bands calculated by Bardayan \textit{et al.} \cite{Bardayan2015} are also included in Fig. \ref{Rate}. However, the calculated uncertainties in the rate considered only known levels and an assumed 3/2$^+$ state at $E_x = 6457$ keV based on the best available information at the time. In this study, this state was taken to have a spin-parity of 5/2$^-$ as reported by Laird \textit{et al.} \cite{Laird}.  

Nova nucleosythesis calculations were performed using the Computational Infrastructure for Nuclear Astrophysics \cite{Smith2006} to investigate how the various reaction rates affect the final $^{18}$F abundance. The calculations were carried out assuming a nova explosion on a 1.0 solar mass CO white dwarf, as well as 1.15, 1.25, and 1.35 solar mass ONeMg white dwarfs. The final $^{18}$F abundance after the explosion was compared for each calculated rate shown in Fig. \ref{Rate}. Rates for the other nuclear reactions were taken from the REACLIB v2.0 library \cite{Cyburt2010}. Isotopic abundances were tracked from $^1$H to $^{54}$Cr in radial zones (27 for CO, 23 for ONeMg) of varying temperature and density calculated from 1D hydrodynamic model calculations \cite{Starrfield1998}. It was found that for increasing white dwarf masses, the final $^{18}$F abundance range was reduced by factors of 2.5, 2.6, 2.5, and 2.4, respectively. Therefore, the range of the maximum detection radius is decreased by a factor of 3.3 by constraining the energies of the 3/2$^+$ states.  This also reduces the average uncertainty on the nova detection probability, which is proportional to the volume of space that can be surveyed by a telescope with minimum $\gamma$-ray flux requirements, by a factor of 2.1.

To summarize, the unknown positions of 3/2$^+$ states in $^{19}$Ne near the proton threshold were a significant source of uncertainty in the astrophysically-important $^{18}$F($p$,$\alpha$)$^{15}$O reaction rate. The previous lack of knowledge regarding the energies of these states resulted in the rate being uncertain by factors of 3 to 33. To search for these levels, triton-$\gamma$-$\gamma$ coincidences from the $^{19}$F($^3$He,$t\gamma$)$^{19}$Ne reaction were measured using GODDESS, and these data constitute the first published result from the GODDESS campaign. An 11/2$^+$ state thought to be at 6440~keV \cite{Laird} was found subthreshold, and a mirror connection was made with the 6500-keV $^{19}$F level based on similar $\gamma$-decay patterns. Decays from levels at $6423\pm3$ and $6441\pm3$~keV to low-spin states provide the first evidence of the expected 3/2$^+$ states, based on the known levels in the mirror nucleus, $^{19}$F. Constraining the level energies reduces the upper limit of the $^{18}$F($p$,$\alpha$)$^{15}$O rate by a factor of 1.5 to 17. Nova nucleosynthesis calculations show the nova detection probability uncertainty is reduced a factor of 2.1 on average. While only upper limits for the flux of the annihilation radiation have been placed by the $\gamma$-ray telescope INTEGRAL (e.g. Siegert \textit{et al.} \cite{Siegert2018}), the e-ASTROGAM project is planned to have a wide field of view and projected to be up to 100 times more sensitive than INTEGRAL \cite{DEANGELIS20181}. Since much of the difficulty in detecting the annihilation radiation from novae is due to the maximum flux occurring prior to the visual maximum, future telescopes with a wide field of view and increased sensitivity will provide the best chance of observation.

\begin{acknowledgments}
The authors thank the ATLAS staff for their efforts during the GODDESS campaign and acknowledge useful discussions with A. M. Laird.  This research was supported in part by the National Science Foundation Grant Number PHY-1419765, the National Nuclear Security Administration under the Stewardship Science Academic Alliances program through DOE Cooperative Agreement DE-NA002132. The authors also acknowledge support from the DOE Office of Science, Office of Nuclear Physics, under contract numbers DE-AC05-00OR22725, DE-FG02-96ER40963, DE-FG02-96ER40978, and Argonne National Laboratory contract number DE-AC02-06CH11357.
\end{acknowledgments}

\bibstyle{apsrev4}

\end{document}